\documentstyle[12pt]{article}

\begin{document}
\title{The search for new physics by the measurement of the 4-jet cross 
section at LHC and FNAL.}
\author{S.I.Bityukov \\ IHEP, Protvino RU-142284, Russia  \\ and \\
N.V.Krasnikov 
\\INR, Moscow 117312, Russia}
\date{April 1997}
\maketitle
\begin{abstract}
We investigate the possibility to look for new physics by the measurement 
of the 4-jet cross section at LHC and FNAL. In particular, we consider the 
model with scalar colour octet and the supersymmetric model with R-parity 
violation. In both models pair produced new particles decay  into 2 jets 
thus leading to 4-jet events. Therefore, the measurement of the distributions 
of 4-jet differential cross section on on the invariant dijet masses allows 
to look for new physics. The main background comes from standard QCD 4-jet 
events. We find that at LHC it would be possible to 
discover scalar colour octet particles with a mass $\leq 900 $ Gev and for 
FNAL the corresponding bound is 175 Gev.    
 
\end{abstract}

\newpage

\section{Introduction}

Many extensions of the standard model predict the existence of new massive 
objects that couple to quarks and gluons, and result in resonant structures 
in the two-jet mass spectrum.  Usually such new objects (excited quarks, 
axigluons, colour octet technirhos, new gauge bosons $W^{'}$, $Z^{'}$, scalar 
diquarks, etc.) are singly produced in hadron-hadron interactions. So such 
new particle will give new resonant type contribution to the QCD two-jet 
cross section. In ref.\cite{1} experimental bounds on new particles decaying 
to dijets have been obtained. However there are examples of new particles 
(scalar colour octets, squarks in model with R-parity violation etc.) 
that produced mainly in pairs and decay to dijets so 
their decays lead to the resonant 
structure for the four-jet differential cross section. The main background 
comes from the standard QCD 4-jet events. Therefore it is very interesting to 
answer the question: is it possible to discover new physics at hadron-hadron 
colliders by the measurement of the 4-jet differential cross section?

In this paper we study the possibility to search for scalar colour 
octets and squarks with R-parity violation by the measurement 
of the 4-jet cross section at LHC and FNAL. We find that at LHC it would 
be possible to discover the scalar colour octets with the mass $\leq 
900 $ Gev and at FNAL the corresponding bound is 175  Gev. 
Analogous bounds take place for the squarks in model with R-parity violation.

In section 2 we describe the phenomenology of scalar octets. In section 3 we 
describe the model with R-parity violation. Section 4 is devoted to the 
discussion of QCD background. In section 5 we discuss the LHC and FNAL 
discovery potential of new physics by the measurement of the four-jet 
cross section. Section 6 contains concluding remarks.

\section{Phenomenology of scalar colour octets}

The relatively light ($M \leq O(1)Tev$) scalar colour 
octets are predicted in some nonsupersymmetric and supersymmetic GUTs
\cite{2,3}. The phenomenology of light scalar octets has been discussed in 
ref.\cite{4}. 
To be precise, in this paper  we consider colour light scalar octets neutral
under $SU(2) \otimes U(1)$ electroweak gauge group. 
Such particles are 
described by the selfconjugate scalar field $\Phi^{\alpha}_{\beta}(x)$ 
 ($(\Phi^{\alpha}_{\beta}(x))^{+} = \Phi^{\beta}_{\alpha}(x),  \,
\Phi^{\alpha}_{\alpha}(x) = 0 $)  interacting only with gluons. Here
$\alpha = 1,2,3  ; \,\beta = 1,2,3$ are $SU(3)$ indices.  The scalar potential
for the scalar octet field $\Phi^{\alpha}_{\beta}(x)$ has the form 
    
\begin{equation}
V(\Phi) = \frac{M^2}{2}Tr(\Phi^{2}) + \frac{\lambda_{1}M}{6}Tr(\Phi^{3}) +
\frac{\lambda_{2}}{12}Tr(\Phi^4)+ \frac{\lambda_{3}}{12}(Tr\Phi^{2})^2
\end{equation}

The term $\frac{\lambda_{1} M}{6}Tr(\Phi^3)$ in the scalar potential (1) 
breaks the discrete symmetry $\Phi \rightarrow -\Phi$. The existence of such 
term in the lagrangian leads to the decay of the scalar octet mainly into 
two gluons through one-loop diagrams similar to the corresponding one-loop 
diagrams describing the Higgs boson decay into two photons. One can find that 
the decay width of the scalar octet is determined by the formula \cite{4}   
\begin{equation}
\Gamma(\Phi \rightarrow g\,g) = \frac{15}{4096\pi^{3}}\alpha_{s}^{2}c^2
\lambda_{1}^{2}M \,,
\end{equation}
where
 
\begin{equation}
c = \int_{0}^{1} \int_{0}^{1-w} \frac{wu}{1-u-w}\,du\,dw = 0.048
\end{equation}
and $\alpha_{s}$ is the effective strong coupling constant at some
normalization point $\mu \sim M $.
Numerically for $\alpha_{s} = 0.12$ we find that 
\begin{equation}
\Gamma(\Phi \rightarrow gg) = 0.39\cdot 10^{-8}\lambda_{1}^{2}M
\end{equation}

>From the requirements that colour $SU(3)$ symmetry is unbroken (the minimum 
$\Phi^{\alpha}_{\beta}(x) = 0 $ is the deepest one) and the effective 
coupling constants $\overline{\lambda}_{2}$ , $\overline{\lambda}_{3}$ 
don't have Landau pole singularities up to the energy $M_{0} = 100\cdot M$ 
we find that $\lambda_{1} \leq O(1)$ . Therefore the decay width of 
the scalar coloured octet is less than $O(10)ev, O(100)ev, O(1)Kev,
O(10)Kev$ for the octet masses $M = 1,\, 10,\, 100,\, 1000$ Gev 
correspondingly. It means that new hadrons composed from scalar octet 
$\Phi$, quarks and gluons ($\overline{q} \Phi q, \,\Phi g, \,qqq\Phi $) are 
longlived even for very high scalar octet mass. Other consequence of the 
smallness of the scalar octet decay width into gluons  is that the cross 
section of the scalar octet single production is exteremely small and that 
the search for the scalar octets in dijet events is hopeless.

Consider pair production of scalar octets at FNAL and LHC. The 
corresponding lowest order formulae for the parton cross sections 
have the form

\begin {equation}
\frac{d\sigma}{dt}(\overline{q}q \rightarrow \Phi \, \Phi) =
\frac{4\pi\alpha^{2}_{s}}{3s^4}(tu - M^4) \,,
\end{equation}
\begin{eqnarray}
\frac{d\sigma}{dt}(gg \rightarrow \Phi \Phi) &=& \frac{\pi\alpha_{s}^{2}}
{s^2}(\frac{27}{32} + \frac{9(u-t)^2}{32s^2})(1 + \frac{2M^2}{u-M^2} +
\frac{2M^2}{t-M^2} +  \\ \nonumber
&&\frac{2M^4}{(u-M^2)^2} + \frac{2M^4}{(t-M^2)^2} +
\frac{4M^4}{(t-M^2)(u-M^2)})\,,
\end{eqnarray}
\begin{equation}
\sigma(\overline{q}q \rightarrow \Phi \Phi) = \frac{2\pi\alpha^{2}_{s}}{9s}
k^{3}\,,
\end{equation}
\begin{equation}
\sigma(gg \rightarrow \Phi \Phi) = \frac{\pi\alpha^{2}_{s}}{s}(\frac{15k}{16}
+\frac{51kM^2}{8s} + \frac{9M^{2}}{2s^2}(s-M^{2})ln(\frac{1-k}{1+k})) \,,
\end{equation} 
where $k = (1 - \frac{4M^{2}}{s})^{\frac{1}{2}}$. We have calculated the 
cross sections for the production of scalar octets at FNAL and LHC using 
the results for the parton distributions of ref.\cite{5}, namely, 
in our calculations we  have used set 1 of the parton distributions of ref. 
\cite{5} at the renormalization point $\mu = 2M$. 
We have found that at LHC the main contribution 
($\geq $ 95\%) comes from the gluon annihilation into two scalar octets 
$gg \rightarrow \Phi \Phi $, whereas at FNAL gluon-gluon and 
quark-antiquark annihilation cross sections are comparable.
The results of our calculations for the 
total cross sections are presented in tables 1, 2 and in figures 1, 3 .
 
For  light scalar octets two gluon and quark-antiquark annihilations 
into two scalar octets give additional contribution to the two-jet 
cross section. However, this additional contribution is rather small. 
For instance, the cross section for gluon-gluon scattering is \cite{6} 
\begin{equation}
\frac{d\sigma}{dt}(gg \rightarrow gg) = \frac{9\pi\alpha^{2}_{s}}{2s^2}
[3 - \frac{tu}{s^2} - \frac{su}{t^2} - \frac{st}{u^2}]
\end{equation} 
Even for the most favorable case $t=u=-\frac{s}{2}$  the cross section
(12) is 20 times less than gluon-gluon cross section (15). So the perspective 
to detect light scalar octets by the measurement of the two-jet cross sections 
looks hopeless. 
For rather big values of the scalar octet mass ($M \geq
O(50)Gev$) the scalar octets decay into two gluons that leads to the 
four-jet events. 

\section{Squarks in model with R-parity violation}

The minimal supersymmetric standard model(MSSM) is considered as a leading 
candidate for supersymmetric generalization of Standard Model \cite{7}. 
In MSSM 
additional symmetry, called R-parity has to be imposed in order to avoid 
renormalizable interactions which violate lepton and baryon numbers. However 
the conservation of R-parity is an ad hoc postulate without deep theoretical 
justification. 

The most general renormalizable R-violating superpotential 
using only MSSM superfields is \cite{8}
\begin{equation}
W = \lambda^{k}_{ij}L_iL_j\bar{E}_k + \lambda_{ijk,1}L_iQ_j\bar{D}_k +
\lambda^{i}_{jk,2}\bar{U}_i\bar{D}_j\bar{D}_k \,.
\end{equation}
Here  i,j,k are generation indices. The couplings $\lambda^{k}_{ij}$ 
are antisymmetric in flavour, $\lambda^{k}_{ij} = -\lambda^{k}_{ji}$. 
Similarly, $\lambda^{i}_{jk,2} =-\lambda^{i}_{kj,2}$. There are 36 lepton 
number nonconserving couplings and 9 baryon number non-conserving couplings 
$\lambda^{i}_{jk,2}$. To avoid rapid proton decay it is necessary to 
put $\lambda^{k}_{ij} = \lambda_{kij,1}=0$ or to put $\lambda^{i}_{jk,2} = 0$.

In this section we consider R-parity violating model with 
$\lambda^{k}_{ij,2}$ different from zero. The constraints on 
$\lambda^{k}_{ij,2}$ couplings have been discussed in refs. 
\cite{9}. It should 
be noted that existing bounds on the R-parity 
violating couplings $\lambda^{t}_{sd,2}, \lambda^{c}_{bd,2}, 
\lambda^{u}_{bs}$ depend on some unknown soft supersymmetry breaking 
parameters of the theory and are not very stringent. The existence of 
R-parity violating interaction  (10) leads to the decay of righthanded 
squarks into two antiquarks, so each of pair produced righthanded squarks will 
decay into two jets resulting in 4-jet signature. We  suppose for 
simplicity that the parameters of the model are such that the branching ratios 
of right-handed squarks to two antiquarks are closed to unit. So the typical 
signature for such scenario is the existence of additional 4-jet events 
arising due to squarks decays into two jets. We have calculated the squark 
cross section in the assumption that gluino mass is much heavier than the 
righthanded squarks and all righthanded squarks are degenerated in mass. 
In our calculations we have used ISASUSY program \cite{10}. The results of our 
calculations are presented in tables 3, 4 and in figures 2, 4 . 
Note that due to nonzero 
R-parity violating interaction we shall have the single squark production 
at supercolliders. At present we don't interested in the single squark particle 
production (it is possible to imagine the situation when all 
$\lambda^{k}_{ij,2}$ coupling constants are small so the single squark 
production is negligible and the righthanded squarks are the lightest 
sparticles so they decay only into two antiquarks).     

\section{QCD background estimates}
 
The main background comes from QCD jets. To estimate QCD background we have 
used PYTHIA 5.7 program \cite{11}. We have used standard UA1 definition of 
jet and took the jet cone equal to $R = 0.4$ and $R= 1$. We have used the 
the transverse momentum cut on jets $p_{T0}$  equal to 100 Gev, 150 Gev, 
200 Gev, 300 Gev for LHC and  50 Gev, 100 Gev for FNAL. We selected 
4-jet events 
such that the invariant dijet masses fulfiled  the conditions 
\begin{equation}
|M_{ij,jet} -M| \leq \delta  ,
\end{equation}
\begin{equation}
|M_{kl,jet} - M| \leq \delta ,
\end{equation}
and moreover the jets have to satisfy the conditions: 

a. $p_{Tjet} \geq p_{T0}$

b. $|\eta_{jet}| \leq \eta_{0}$

Here  $i,j,k,l =1,2,3,4$ label the jet number 
and $i \neq j, i \neq k, i \neq l, j \neq k, j \neq l,
k \neq i, k \neq j, k \neq l$. For LHC we took $\eta_{0} =2.5$ and for 
FNAL we have used $\eta_{0} = 0.5$.   
The parameter $\delta$ determines the accuracy 
of the dijet invariant mass determination. In our analysis for LHC we  have 
used 
$\delta = 50$ Gev (optimistic variant) and $\delta = 100$ Gev 
(realistic variant).
 For FNAL we have used $\delta = 25$ Gev. Both  
CMS and ATLAS detectors will be able to measure the dijet invariant mass 
with the accuracy 10 percent or even better for the case of big invariant 
dijet masses \cite{12,13}. Our realistic variant for $\delta$ corresponds to  
approximately 10 percent dijet invariant mass resolution and optimistic 
variant corresponds approximately to 5 percent accuracy in the dijet mass 
determination. For FNAL $\delta = 25$ Gev corresponds to $\approx$ 10 percent 
dijet mass determination for $M_{dijet} \leq 225$ Gev  
We have generated one million QCD events for each 
value of $P_{T0}$ and $R$ to find 4-jet QCD background satisfying mass 
cuts (11,12) both for LHC and FNAL. As for FNAL typical accuracy in the 
determination of dijet invariant mass is 10 percent \cite{1}. The results of 
 our  QCD background calculations are presented in tables 5 - 11 and in  
figures 5 -10 . In our calculations we took the LHC total luminosity 
equal to $L_t = 10^{4} pb^{-1}$ and for FNAL we have used $L_{t} = 
100 pb^{-1}$. In tables 5 - 11 $\sigma(4 jets, back.)$ denotes the QCD 
background 4-jet cross section satisfying the conditions (11,12) 
and with the cut on $p_{T}$ and $\eta$. For LHC we have used 
in tables 5 - 9  $\delta = 100$ Gev (realistic variant). For FNAL we have 
used $\delta = 25$ Gev ($\approx$ 10 percent dijet mass accuracy 
deternination). For instance, the value 
$\sigma (4 jets, back.)$ for the $M = 0.3$ Tev in table 5 means that both 
dijet invariant masses have to be between (300 - 100) Gev and (300 + 100) Gev. 
The value $\sigma(sig.disc.)$ means the lower value of the acceptable 
cross section corresponding to new physics which can be detected at 
$5 \sigma$ level. The value $\sigma^{ac}(\Phi \Phi)$ denotes the cross section 
$\sigma( pp \rightarrow (\Phi \rightarrow jet1 + jet2) + (\Phi \rightarrow 
jet3 + jet4) + ...)$ , where jets 1,2,3,4 satisfy the $p_{T}$ and $\eta$ cuts. 
The value $\sigma^{ac}(squarks)$ has the similar meaning.

\section{LHC and FNAL discovery potentials}

As it has been mentioned before in our concrete calculations 
we take the total LHC luminosity equal to 
$L_{t,LHC} = 10^{4}pb^{-1}$ and the total FNAL luminosity $L_{t,FNAL} =
100 pb^{-1}$. According to standard folklore (unfortunately 
folklore $\neq$ theorem statement) we suppose that new physics will be 
discovered by the measurement of the 4-jet events provided that
 
\begin{equation}
Significance = \frac{N_{signal}}{\sqrt{N_{background}}} \geq 5 ,
\end{equation}
where $N_{signal} = \sigma_{signal}L_{t}$ and $N_{background} = 
\sigma_{background}L_{t}$. 
The results of our calculations are presented in tables 5 - 11 and in 
figures 11 - 15.
It appears that the most promising cut for the 
search for scalar octets at LHC corresponds to $p_{T0} = 300$ Gev and the 
jet definition with the cone equal to $R=0.4$ (table 8). As it follows from 
the table 8 it is possible to discover at $5 \sigma $ level scalar octets with 
the mass up to 900 Gev and the squarks in the model with R-parity violation 
with the mass up to 1100 Gev. As for FNAL the most promising cut corresponds 
to $p_{T0} = 100$ Gev ( Table 12).  As it follows from the table 12 FNAL is 
able to discover scalar octets and squarks with the masses lighter than 
175 Gev. As we know up to now there was not analysis of 
dijet mass distributions for 4-jet events both in CDF and D0. It would be very 
intersting to perform such analysis.  In our estimates we used PYTHIA 4 -jets 
estimates. It is well known that PYTHIA underestimates the number of many jets 
events. Suppose that the real 4-jets background is 10 times bigger than the 
PYTHIA background. In this case for LHC the discovery potential of 
scalar octets and squarks will be for the masses up to $\approx 700$ Gev and 
for FNAL up to $\approx 150$ Gev. However the increase of luminosity 
for LHC up to $L_{t} = 10^{5} pb^{-1}$ (the luminosity for a year 
after 2 first years of exploitation) and for FNAL up to $L_{t} = 1000 pb^{-1}$ 
(upgraded TEVATRON) just will compensate the factor 10 in background 
cross section.  More reliable estimates of the 4-jet cross section are 
necessary.

\section{Conclusion}

To conclude, in this paper we have studied the perspectives of the discovery
of scalar octets and squarks in the model with R-parity violation
at FNAL and LHC. We have found that  
scalar octets and squarks could be discovered at LHC and FNAL by the 
measurement of the distributions of the differential 4-jet cross section 
on the invariant two-jet masses with the mass up to 900 - 1100 Gev(LHC) and 
175 - 200 Gev(FNAL). One of the main problems here is an accurate estimate 
of QCD background. We have used PYTHIA to estimate QCD background. In general 
PYTHIA gives the 4 jet cross section with the accuracy of factor 2 -5. 
Therefore more careful calculation of QCD background is necessary. 
Nevertheless our results are very encouraging. Moreover, it is necessary to 
estimate more carefully the accuracy of the dijet invariant mass determination 
at CMS and ATLAS. However we think that our numbers  5 percent and 10 percent 
(optimistic  and realistic variants) for the 
estimation of the dijet invariant mass accuracy determination are reasonable.   
     
We are  indebted to the participants of Daniel Denegri seminar on physics 
simulations for useful discussions.  
We thank Igor Semenuk for the help. 
 
The research described in this publication was made possible in part by 
Award No RP1-187 of the U.S. Civilian Research and Development Foundation for 
the Independent States of the Former Soviet Union(CRDF).

\newpage

Table 1. The cross section $\sigma(p p \rightarrow \Phi \Phi
\, + \, ...)$ in pb for different values of octet masses and 
at LHC.

\begin{center}
\begin{tabular}{|l|l|l|l|l|l|l|l|}
\hline
M(Tev) & 0.2 & 0.3 & 0.4 & 0.5 & 0.7 & 0.9 & 1.1 \\ 
\hline
$\sigma$ &701  & 84 &20 &7.4 & 1.1 & 0.18 &0.055  \\
\hline
\end{tabular}
\end{center}

Table 2. The cross section $\sigma(\overline{p} p \rightarrow \Phi \Phi \, +
\, ...)$ in pb for different values of octet masses  
at FNAL.

\begin{center}
\begin{tabular}{|l|l|l|l|l|l|l|l|l|l|}
\hline
M(Gev) &  125 & 150 & 175 & 200 & 225 & 250 & 275 & 300&325 \\
\hline
$\sigma$  & 11 & 3.6 & 1.1 &  0.42 &0.21 & 0.074 & 0.030&0.014&0.0067   \\
\hline
\end{tabular}
\end{center}

Table 3. The cross sections for the production of 6 mass degenerate 
righthanded squarks in pb at LHC for the case of very heavy gluino.

\begin{center}
\begin{tabular}{|l|l|l|l|l|l|l|l|l|l|}
\hline
M(Tev) &0.2& 0.3&0.4&  0.5 &0.6& 0.7 &0.8& 0.9 & 1.1\\
\hline
$\sigma$  &300& 56&14 & 4.7&2.1 & 0.81 &0.47& 0.24 & 0.074 \\
\hline 
\end{tabular}
\end{center}

Table 4. The cross sections for the production of of 6 mass degenerate 
righthanded squarks in pb at FNAL for the case of very heavy gluino.
\begin{center}
\begin{tabular}{|l|l|l|l|l|l|l|l|l|l|}
\hline
M(Gev) &125&150&175&200&225&250&275&300&325 \\
\hline
$\sigma$  &8.2&3.8&1.7&0.72&0.37&0.17&0.083&0.036&0.017 \\
\hline
\end{tabular}
\end{center}

Table 5. The background and signal acceptance cross sections for 
LHC ($p_{T} \geq p_{T0} = 100$ Gev, $R =0.4$, $L_t = 10^{4}pb^{-1}$).
All cross sections are in pb.

\begin{center}
\begin{tabular}{|l|l|l|l|l|l| }
\hline
M(Tev) & 0.3 & 0.5 & 0.7 & 0.9 & 1.1 \\
\hline
$\sigma(4 jets,back.)$ & 1200 & 430 & 84 & 19 & 3.8 \\
\hline
$\sigma(sig. disc.)$ & 1.7 & 1.1 & 0.46 & 0.22 &0.1 \\
\hline
$\sigma^{ac}(\Phi \Phi)$ & 25 & 3.7 & 0.61 & 0.12 & 0.040 \\
\hline
$\sigma^{ac}(squarks)$ & 17& 2.3 & 0.44 & 0.13 & 0.053 \\
\hline
\end{tabular}
\end{center}
 
Table 6.
The background and signal acceptance cross sections for LHC 
( $p_{T} \geq p_{T0} = 150$ Gev, $R = 1$, $L_t = 10^{4}pb^{-1}$). 
All cross sections are in pb.

\begin{center}
\begin{tabular}{|l|l|l|l|l|l|l|}
\hline
M(Tev) & 0.3 & 0.5 & 0.7 & 0.9 & 1.1  \\
\hline
$\sigma(4 jets, back.)$ & 58 & 165 & 54 & 22 & 4.8 \\
\hline
$\sigma(sig. disc.)$ & 0.38 & 0.64 & 0.37 &  0.23 & 0.11  \\
\hline
$\sigma^{ac}(\Phi \Phi)$ & 18 & 2.2 & 0.50 & 0.090 & 0.037  \\
\hline
$\sigma^{ac}(squarks)$ & 12 & 1.4 & 0.36 & 0.12 & 0.050 \\
\hline
\end{tabular}
\end{center}

Table 7. The background and signal acceptance cross sections for LHC 
($p_{T} \geq p_{T0} = 200$ Gev, $R = 0.4$, $L_t = 10^{4} pb^{-1}$). All 
cross sections are in pb.

\begin{center}
\begin{tabular}{|l|l|l|l|l|l|l|}
\hline
M(Tev) & 0.3 & 0.5 & 0.7 & 0.9 & 1.1 \\
\hline
$\sigma(4 jets, back.)$ & 4 & 14 & 12 & 4 & 1.8 \\
\hline
$\sigma(sig.disc.)$ & 0.10 & 0.19 & 0.17 & 0.11 & 0.07 \\
\hline
$\sigma^{ac}(\Phi \Phi)$ & 12 & 1.9 & 0.44 & 0.081 & 0.028 \\
\hline
$\sigma^{ac}(squarks)$ & 7.8 & 1.2 & 0.32 & 0.11 & 0.038 \\
\hline
\end{tabular}
\end{center}

Table 8. The background and acceptance cross sections at LHC ($p_{T} 
\geq p_{T0} = 300$ Gev, $R = 0.4$, $L_t = 10^{4}pb^{-1}$). All cross sections 
are in pb.

\begin{center}
\begin{tabular}{|l|l|l|l|l|l|}
\hline
M(Tev) & 0.3 & 0.5 & 0.7 & 0.9 & 1.1 \\
\hline
$\sigma(4 jets, back.)$ & 0.22 & 0.13 & 0.73 & 0.90 & 0.34 \\
\hline
$\sigma(sig. disc.)$ & 0.023 & 0.018 & 0.043 & 0.048 &  0.029 \\
\hline
$\sigma^{ac}(\Phi \Phi)$ & 6.7 & 1.2 & 0.33 & 0.068 & 0.024 \\
\hline
$\sigma^{ac}(squarks)$ & 4.5 & 0.74 & 0.24 & 0.092 & 0.033  \\
\hline
\end{tabular}
\end{center}

Table 9. The background and acceptance 4 jet cross sections at LHC 
($p_{T} \geq p_{T0} = 300$ Gev, $R =1$, $L_t =10^{4} pb^{-1}$). All 
cross sections are in pb.

\begin{center}
\begin{tabular}{|l|l|l|l|l|l|l|}
\hline
M(Tev) & 0.3 & 0.5 & 0.7 & 0.9 & 1.1 \\
\hline
$\sigma(4 jets, back.)$ & 0.043 & 0.26 & 0.60 & 1.1 & 0.43 \\
\hline
$\sigma(sig. disc.)$ & 0.011 & 0.026 & 0.040 & 0.052 & 0.033 \\
\hline
$\sigma^{ac}(\Phi \Phi)$ & 6.7 & 1.2 & 0.33 & 0.068 & 0.024 \\
\hline
$\sigma^{ac}(squarks)$ & 4.5 & 0.74 & 0.24 & 0.092 & 0.033 \\
\hline
\end{tabular}
\end{center}

Table 10. The background and accepted 4 jet cross sections at FNAL 
($p_{T} \geq p_{T0} = 50$ Gev, $R= 0.4$, $L_t=10^{2}pb^{-1}$). All cross 
sections are in pb.

\begin{center}
\begin{tabular}{|l|l|l|l|l|}
\hline
M(Gev) & 125 & 175 & 225 & 275 \\
\hline
$\sigma(4 jets back.)$ & 0.34 & 0.51 & 0.17 & 0.17 \\
\hline
$\sigma^{ac}(sig. disc.)$ & 0.29 & 0.36 & 0.21 & $\leq 0.21$ \\
\hline
$\sigma^{ac}(\Phi \Phi)$ & 2.2 & 0.22 & 0.042 & 0.0061 \\
\hline
$\sigma^{ac}(squarks)$ & 1.6 & 0.36 & 0.078 & 0.017 \\  
\hline
\end{tabular}
\end{center}

Table 11. The background and acceptance 4-jet cross sections at 
FNAL ($p_{T} \geq p_{T0} = 100$ Gev, $R =1$, $L_t =10^{2}pb^{-1}$). All 
cross sections are in pb.

\begin{center}
\begin{tabular}{|l|l|l|l|l|l|}
\hline
M(Gev) & 125 & 175 & 225 & 275 \\
$\sigma(4 jets, back.)$ & $\leq 0.01$ & $\leq 0.01$ & $\leq 0.01$ & $\leq 0.01 $ \\
\hline
$\sigma^{ac}(sig. disc.)$ & $\geq 0.05$ & $\geq 0.05 $ & $\geq 0.05$ & $\geq 0.05$ \\
\hline
$\sigma^{ac}(\Phi \Phi)$ & 1.2 & 0.103 & 0.025 & 0.0036 \\
\hline
$\sigma^{ac}(squarks)$ & 0.90 & 0.20 & 0.044 & 0.011 \\
\hline
\end {tabular}
\end{center}

\end{document}